\documentclass[twocolumn,showpacs,aps,prl,superscriptaddress]{revtex4}

\usepackage{graphicx}
\usepackage{dcolumn}
\usepackage{amsmath}
\usepackage{epsfig}
\usepackage{rotating}

\RequirePackage{xspace}

\usepackage{relsize}
\def\babar{\mbox{\slshape B\kern-0.1em{\smaller A}\kern-0.1em
    B\kern-0.1em{\smaller A\kern-0.2em R}}}

\def\epem       {\ensuremath{e^+e^-}\xspace}

\def\ellell     {\ensuremath{\ell^+ \ell^-}\xspace}

\def\qbar  {\ensuremath{\overline q}\xspace}
\def\qqbar {\ensuremath{q\overline q}\xspace}

\def\cbar  {\ensuremath{\overline c}\xspace}

\def\piz   {\ensuremath{\pi^0}\xspace}

\def\pip   {\ensuremath{\pi^+}\xspace}
\def\pim   {\ensuremath{\pi^-}\xspace}
\def\pipi  {\ensuremath{\pi^+\pi^-}\xspace}
\def\pipm  {\ensuremath{\pi^\pm}\xspace}
\def\pimp  {\ensuremath{\pi^\mp}\xspace}

\def\Kbar  {\kern 0.2em\overline{\kern -0.2em K}{}\xspace}

\def\Kz    {\ensuremath{K^0}\xspace}
\def\Kzb   {\ensuremath{\Kbar^0}\xspace}
\def\KzKzb {\ensuremath{\Kz \kern -0.16em \Kzb}\xspace}
\def\Kp    {\ensuremath{K^+}\xspace}
\def\Km    {\ensuremath{K^-}\xspace}
\def\Kpm   {\ensuremath{K^\pm}\xspace}

\def\KpKm  {\ensuremath{\Kp \kern -0.16em \Km}\xspace}
\def\KS    {\ensuremath{K^0_{\scriptscriptstyle S}}\xspace}

\def\Dbar    {\kern 0.2em\overline{\kern -0.2em D}{}\xspace}

\def\Dz      {\ensuremath{D^0}\xspace}
\def\Dzb     {\ensuremath{\Dbar^0}\xspace}
\def\DzDzb   {\ensuremath{\Dz {\kern -0.16em \Dzb}}\xspace}
\def\Dp      {\ensuremath{D^+}\xspace}
\def\Dm      {\ensuremath{D^-}\xspace}

\def\DpDm    {\ensuremath{\Dp {\kern -0.16em \Dm}}\xspace}

\def\B       {\ensuremath{B}\xspace}
\def\Bbar    {\kern 0.18em\overline{\kern -0.18em B}{}\xspace}

\def\BB      {\ensuremath{B\Bbar}\xspace} 
\def\Bz      {\ensuremath{B^0}\xspace}
\def\Bzb     {\ensuremath{\Bbar^0}\xspace}
\def\BzBzb   {\ensuremath{\Bz {\kern -0.16em \Bzb}}\xspace}
\def\Bu      {\ensuremath{B^+}\xspace}
\def\Bub     {\ensuremath{B^-}\xspace}
\def\Bp      {\ensuremath{\Bu}\xspace}
\def\Bm      {\ensuremath{\Bub}\xspace}
\def\Bpm     {\ensuremath{B^\pm}\xspace}

\def\BpBm    {\ensuremath{\Bu {\kern -0.16em \Bub}}\xspace}

\def\BorBbar    {\kern 0.18em\optbar{\kern -0.18em B}{}\xspace}
\def\DorDbar    {\kern 0.18em\optbar{\kern -0.18em D}{}\xspace}
\def\KorKbar    {\kern 0.18em\optbar{\kern -0.18em K}{}\xspace}

\def\jpsi     {\ensuremath{{J\mskip -3mu/\mskip -2mu\psi\mskip 2mu}}\xspace}
\def\psitwos  {\ensuremath{\psi{(2S)}}\xspace}

\def\chiczero {\ensuremath{\chi_{c0}}\xspace}

\def\chictwo  {\ensuremath{\chi_{c2}}\xspace}
\mathchardef\Upsilon="7107
\def\Y#1S{\ensuremath{\Upsilon{(#1S)}}\xspace}

\def\FourS {\Y4S}

\mathchardef\Deltares="7101
\mathchardef\Xi="7104
\mathchardef\Lambda="7103
\mathchardef\Sigma="7106
\mathchardef\Omega="710A

\def\Deltabar{\kern 0.25em\overline{\kern -0.25em \Deltares}{}\xspace}
\def\Lbar{\kern 0.2em\overline{\kern -0.2em\Lambda\kern 0.05em}\kern-0.05em{}\xspace}
\def\Sigbar{\kern 0.2em\overline{\kern -0.2em \Sigma}{}\xspace}
\def\Xibar{\kern 0.2em\overline{\kern -0.2em \Xi}{}\xspace}
\def\Obar{\kern 0.2em\overline{\kern -0.2em \Omega}{}\xspace}
\def\Nbar{\kern 0.2em\overline{\kern -0.2em N}{}\xspace}
\def\Xb{\kern 0.2em\overline{\kern -0.2em X}{}\xspace}

\def\BR         {{\ensuremath{\cal B}\xspace}}

\def\mes        {\mbox{$m_{\rm ES}$}\xspace}

\def\DeltaE     {\mbox{$\Delta E$}\xspace}

\newcommand{\tev}{\ensuremath{\mathrm{\,Te\kern -0.1em V}}\xspace}
\newcommand{\gev}{\ensuremath{\mathrm{\,Ge\kern -0.1em V}}\xspace}
\newcommand{\mev}{\ensuremath{\mathrm{\,Me\kern -0.1em V}}\xspace}
\newcommand{\kev}{\ensuremath{\mathrm{\,ke\kern -0.1em V}}\xspace}
\newcommand{\ev}{\ensuremath{\mathrm{\,e\kern -0.1em V}}\xspace}
\newcommand{\gevc}{\ensuremath{{\mathrm{\,Ge\kern -0.1em V\!/}c}}\xspace}
\newcommand{\mevc}{\ensuremath{{\mathrm{\,Me\kern -0.1em V\!/}c}}\xspace}
\newcommand{\gevcc}{\ensuremath{{\mathrm{\,Ge\kern -0.1em V\!/}c^2}}\xspace}
\newcommand{\mevcc}{\ensuremath{{\mathrm{\,Me\kern -0.1em V\!/}c^2}}\xspace}

\def\cm   {\ensuremath{{\rm \,cm}}\xspace}

\def\invfb   {\ensuremath{\mbox{\,fb}^{-1}}\xspace}

\def\mus  {\ensuremath{\rm \,\mus}\xspace}

\def\mus        {\ensuremath{\,\mu{\rm s}}\xspace}

\def\to                 {\ensuremath{\rightarrow}\xspace}

\def\pep2{PEP-II}

\newcommand{\chisq}{\ensuremath{\chi^2}\xspace}

\def\gsim{{~\raise.15em\hbox{$>$}\kern-.85em
          \lower.35em\hbox{$\sim$}~}\xspace}
\def\lsim{{~\raise.15em\hbox{$<$}\kern-.85em
          \lower.35em\hbox{$\sim$}~}\xspace}

\def\CP                {\ensuremath{C\!P}\xspace}

\xspace

\newcommand{\tabref}[1]{Table~\ref{tab:#1}}

\def\jetset74   {\mbox{\tt Jetset \hspace{-0.5em}7.\hspace{-0.2em}4}\xspace}

\renewcommand{\eqref}[1]{Eq.~(\ref{eq:#1})}

\newcommand{\appref}[1]{Appendix~\ref{sec:#1}}

\newcommand{\onreslumi}  {\mbox{424\invfb}}
\newcommand{\offreslumi} {\mbox{44\invfb}}
\newcommand{\bbpairs}    {\mbox{$(465\pm5)\times10^6$}}

\newcommand{\nbb}        {\mbox{$N_{\BB}$}}

\newcommand{\Lzero}      {\mbox{$L_0$}}
\newcommand{\Ltwo}       {\mbox{$L_2$}}
\newcommand{\LtwoOverLzero} {\mbox{$\Ltwo/\Lzero$}}

\newcommand{\PPP}                {\mbox{$\pipm \pipm \pimp$}}
\newcommand{\KPP}                {\mbox{$\Kpm  \pipm \pimp$}}

\newcommand{\BtoPPP}             {\mbox{$\Bpm \to \PPP$}}
\newcommand{\BtoKPP}             {\mbox{$\Bpm \to \KPP$}}

\newcommand{\PPPz}               {\mbox{$\pip\pim\pi^0$}}
\newcommand{\BtoPPPz}            {\mbox{$\Bz \to \PPPz$}}

\newcommand{\Bztopippim}         {\mbox{$\Bz \to \pi^+\pi^-$}}
\newcommand{\BztoKmpip}          {\mbox{$\Bz \to K^-\pi^+$}}

\newcommand{\rhoz}               {\mbox{$\rho^0$}}

\newcommand{\rhoI}               {\mbox{$\rhoz(770)$}}

\newcommand{\rhoIpipm}           {\mbox{$\rhoI \pipm$}}

\newcommand{\rhoItopippim}       {\mbox{$\rhoI \to \pip\pim$}}

\newcommand{\omegaI}              {\mbox{$\omega(782)$}}

\newcommand{\omegaIpipm}          {\mbox{$\omegaI \pipm$}}

\newcommand{\BtoomegaIpi}         {\mbox{$\Bpm \to \omegaIpipm$}}

\newcommand{\fz}                 {\mbox{$f_0$}}
\newcommand{\fI}                 {\mbox{$\fz(980)$}}

\newcommand{\fIpipm}             {\mbox{$\fI \pipm$}}

\newcommand{\fItopippim}         {\mbox{$\fI \to \pip\pim$}}

\newcommand{\rhoII}              {\mbox{$\rhoz(1450)$}}

\newcommand{\rhoIIpipm}            {\mbox{$\rhoII \pipm$}}

\newcommand{\rhoIItopippim}       {\mbox{$\rhoII \to \pip\pim$}}

\newcommand{\fII}                {\mbox{$f_2(1270)$}}

\newcommand{\fIIpipm}             {\mbox{$\fII \pipm$}}

\newcommand{\fIItopippim}        {\mbox{$\fII \to \pip\pim$}}

\newcommand{\fIII}               {\mbox{$f_0(1370)$}}

\newcommand{\fIIIpipm}            {\mbox{$\fIII \pipm$}}

\newcommand{\fIIItopippim}       {\mbox{$\fIII \to \pip\pim$}}

\newcommand{\chiczpipm}          {\mbox{$\chiczero \pipm$}}

\newcommand{\Bpmtochiczpipm}     {\mbox{$\Bpm \to \chiczpipm$}}
\newcommand{\chicztopippim}      {\mbox{$\chiczero \to \pip\pim$}}

\newcommand{\chictwotopippim}    {\mbox{$\chictwo \to \pip\pim$}}
\newcommand{\chictwopipm}        {\mbox{$\chictwo \pipm$}}

\newcommand{\Dzbpip}             {\mbox{$\Dzb \pip$}}
\newcommand{\BptoDzbpip}         {\mbox{$\Bp \to \Dzbpip$}}

\newcommand{\DzbtoKpi}           {\mbox{$\Dzb \to \Kp\pim$}}

\newcommand{\JPsitoll}           {\mbox{$\jpsi \to \ellell$}}

\newcommand{\Psitoll}            {\mbox{$\psitwos \to \ellell$}}

\def\CalACP {{\ensuremath{{\cal A}_{\CP}}\xspace}}

\def\etalc              {{\it et~al.~}}

\newcommand{\occ}[1]{\multicolumn{1}{c}{#1}}
\newcommand{\tcc}[1]{\multicolumn{2}{c}{#1}}
\newcommand{\fcc}[1]{\multicolumn{4}{c}{#1}}

\newcommand{\Dx}{\mbox{$\Delta x$}}
\newcommand{\Dy}{\mbox{$\Delta y$}}

\newcommand{\shortfigref}[1]{Fig.~\ref{fig:#1}}

\def\Abar {\kern 0.2em\overline{\kern -0.2em A}{}\xspace}
\def\Fbar {\kern 0.2em\overline{\kern -0.2em F}{}\xspace}
\def\cbar {\kern 0.08em\overline{\kern -0.08em c}{}\xspace}
\def\qbar {\kern 0.1em\overline{\kern -0.1em q}{}\xspace}
\def\Qbar {\kern 0.2em\overline{\kern -0.2em Q}{}\xspace}
\def\epsilonbar {\kern 0.1em\overline{\kern -0.1em \varepsilon}{}\xspace}

\newcommand{\BABARPubYear}    {08}
\newcommand{\BABARPubNumber}  {054}
\newcommand{\SLACPubNumber} {13494}

\newcommand{\gevccSq}{\ensuremath{{\mathrm{\,Ge\kern -0.1em V^2\!/}c^4}}\xspace}
\newcommand{\gevccSqinv}{\ensuremath{{\mathrm{\,Ge\kern -0.1em V^{-2}}c^4}}\xspace}

\begin{document}

\begin{flushleft}
  SLAC-PUB-\SLACPubNumber \\
  \babar-PUB-\BABARPubYear/\BABARPubNumber
\end{flushleft}

\title{
  {
    \large \bf \boldmath 
    Dalitz Plot Analysis of $\Bpm\to\PPP$ Decays
  }
}

\author{B.~Aubert}
\author{Y.~Karyotakis}
\author{J.~P.~Lees}
\author{V.~Poireau}
\author{E.~Prencipe}
\author{X.~Prudent}
\author{V.~Tisserand}
\affiliation{Laboratoire d'Annecy-le-Vieux de Physique des Particules (LAPP), Université de Savoie, CNRS/IN2P3, F-74941 Annecy-Le-Vieux, France }
\author{J.~Garra~Tico}
\author{E.~Grauges}
\affiliation{Universitat de Barcelona, Facultat de Fisica, Departament ECM, E-08028 Barcelona, Spain }
\author{L.~Lopez$^{ab}$ }
\author{A.~Palano$^{ab}$ }
\author{M.~Pappagallo$^{ab}$ }
\affiliation{INFN Sezione di Bari$^{a}$; Dipartmento di Fisica, Universit\`a di Bari$^{b}$, I-70126 Bari, Italy }
% ...
\collaboration{The \babar\ Collaboration}
\noaffiliation

\date{
  \today
}

\begin{abstract} 
\noindent

We present a Dalitz plot analysis of charmless \Bpm\ decays
to the final state \PPP\ using a sample of \bbpairs\ \BB\ pairs 
collected by the \babar\ experiment at $\sqrt{s}=10.58\gev$.
We measure the branching fractions 
$\BR(\Bpm \to \PPP) = (15.2\pm0.6\pm1.2\pm0.4)\times 10^{-6}$,
$\BR(\Bpm \to \rhoIpipm) = (8.1\pm0.7\pm1.2^{+0.4}_{-1.1})\times 10^{-6}$,
$\BR(\Bpm \to \fIIpipm) = (1.57\pm0.42\pm0.16\,^{+0.53}_{-0.19})\times 10^{-6}$, and
$\BR(\Bpm \to \PPP\ {\rm{nonresonant}}) = (5.3\pm0.7\pm0.6^{+1.1}_{-0.5})\times 10^{-6}$,
where the uncertainties are statistical, systematic, and model-dependent, respectively.
Measurements of branching fractions for the modes $\Bpm\to\rhoIIpipm$ and $\Bpm\to\fIIIpipm$
are also presented. We observe no significant direct \CP\ asymmetries for the above modes, and
there is no evidence for the decays $\Bpm\to\fIpipm$, $\Bpm\to\chiczpipm$, or $\Bpm\to\chictwopipm$.

\end{abstract}

\maketitle

\section{Introduction}

Decays of \B\ mesons to three-body charmless final states probe the
properties of the weak interaction through their dependence on the
complex quark couplings described in the Cabibbo-Kobayashi-Maskawa (CKM) 
matrix~\cite{Cabibbo:1963yz,Kobayashi:1973fv}. Furthermore, these
decays test dynamical models for hadronic \B\ decays.

One can measure direct \CP\ asymmetries and constrain magnitudes and
phases of the CKM matrix elements using individual channels that appear
as intermediate resonances in the $\BtoPPP$ decay.
For example, the CKM angle $\gamma$ could be extracted from the
interference between the decay \Bpmtochiczpipm, which has no \CP-violating
phase (in the standard parametrization), and other modes such as
\Bpm\to\rhoIpipm~\cite{Eilam:1995nz,Deshpande:1995nu,Bediaga:1998ma,Bajc:1998bs,Deandrea:2000zg,Blanco:2000gw}.

Studies of $\BtoPPP$ can also be useful for a precise
measurement of the CKM angle $\alpha$. A theoretically clean determination 
of this angle can be obtained from the decay-time dependence
of the interference between $\Bz\to\rho^+\pi^-$, $\Bz\to\rho^-\pi^+$,
and $\Bz\to\rho^0\pi^0$ via the analysis of the Dalitz plot for
$\Bz\to\pip\pim\piz$ decays~\cite{Snyder:1993mx} (recently implemented
by \babar~\cite{Aubert:2007jn} and Belle~\cite{Kusaka:2007dv,Kusaka:2007mj}). 
Charged \B\ decays offer a large statistics sample with which to determine
additional resonant or nonresonant contributions to the three-pion Dalitz
plot that can affect the measurement of $\alpha$.
For example, the Dalitz plot analysis of $\BtoPPP$ allows one to check for effects
from \BtoomegaIpi, that could cause large direct \CP\ violation due to
$\rho$--$\omega$ mixing~\cite{Leitner:2002xh}. 
It is particularly important to limit the possible effects of broad scalar
structures [including the so-called $f_0(600)$ or $\sigma$] and nonresonant
contributions~\cite{Deandrea:2000ce,Gardner:2001gc,Tandean:2002pe,Cheng:2002qu,Meissner:2003pd,Cheng:2007si}. 

Furthermore, a number of unexplained structures have been observed in
charmless $B$ decays to
$K\pi\pi$~\cite{Aubert:2005ce,Aubert:2008bj,Aubert:2007vi,Garmash:2004wa,Garmash:2005rv,Garmash:2006fh},
$KK\pi$~\cite{Aubert:2007xb,Aubert:2008aw}, and
$KKK$~\cite{Garmash:2004wa,Aubert:2006nu,Aubert:2007sd} final states.
Verifying the presence of these structures in $\BtoPPP$ decays would help
to determine their nature and involvement in hadronic $B$ decays.

In this paper we present an amplitude analysis of
\BtoPPP\ decays based on a \onreslumi\ data sample containing
\bbpairs\ \BB\ pairs (\nbb).
The data were collected with the \babar\ detector~\cite{Aubert:2001tu} at the
\pep2\ asymmetric-energy \epem\ storage rings~\cite{Kozanecki:2000cm}
operating at the $\FourS$ resonance with center-of-mass (CM) energy of
$\sqrt{s}=10.58\gev$.
An additional total integrated luminosity of \offreslumi\ was recorded
$40\mev$ below the $\FourS$ resonance (``off-peak'' data) and was used to
study backgrounds.
Compared to our previous publication~\cite{Aubert:2005sk}, 
in addition to doubling the data sample we have
included several improvements in reconstruction algorithms 
that enhance the signal efficiency,
made numerous modifications to the analysis to increase the sensitivity
to direct \CP\ violation effects (for example, 
by including more discriminating variables in the maximum likelihood fit),
and improved our model of the Dalitz plot structure.

The remainder of the paper is organized as follows: Sec. II describes
the amplitude analysis formalism, Secs. III and IV give details about the 
selection of signal $B$ decays and how backgrounds are considered, Sec. V
presents the results from the likelihood fit, Sec. VI gives an account
of the various sources of systematic uncertainties, while Sec. VII summarizes
the results.

\section{Amplitude Analysis Formalism}

A number of intermediate states contribute to the decay \BtoPPP.
We determine their contributions with a maximum likelihood fit to the
distribution of events in the Dalitz plot. This procedure has been described
in detail in our previous 
publications~\cite{Aubert:2005sk,Aubert:2005ce,Aubert:2008bj}.

The \BtoPPP\ decay contains two same-sign pions in the final state. We
distinguish these particles according to the invariant mass they make when
combined with the oppositely charged pion, and draw the Dalitz plot in terms
of heavy and light invariant masses-squared of the $\pi^{\pm}\pi^{\mp}$ systems 
(denoted $m_{\rm max}^2$ and $m_{\rm min}^2$, respectively), so that each
candidate has a uniquely defined position. Moreover, we explicitly enforce
the symmetrization of the total amplitude under exchange of identical bosons.

The total signal amplitudes for \Bp\ and \Bm\ decays are given by
\begin{eqnarray}
\label{eq:totAmp}
  A \equiv A(m_{\rm max}^2,m_{\rm min}^2) &=& 
  \sum_j c_j F_j(m_{\rm max}^2,m_{\rm min}^2) \,,\\ \nonumber
  \Abar \equiv \Abar (m_{\rm max}^2,m_{\rm min}^2) &=& 
  \sum_j \cbar_j \Fbar_j(m_{\rm max}^2,m_{\rm min}^2) \,.
\end{eqnarray}
The complex coefficients $c_j$ and $\cbar_j$ for a given decay mode $j$
contain all the weak phase dependence. Since the $F_j$ terms contain only strong dynamics, 
$F_j \equiv \Fbar_j$. 
We use the following parametrization~\cite{Aubert:2008bj} for the amplitude coefficients:
\begin{eqnarray}
c_j &=& (x_j + \Delta x_j) + i (y_j + \Delta y_j) \\ \nonumber
\cbar_j &=& (x_j - \Delta x_j) + i (y_j - \Delta y_j) \, .
\label{eq:cjDef}
\end{eqnarray}
In this approach, $x_j$ and $y_j$ ($\Delta x_j$ and $\Delta y_j$) are the \CP-conserving
(-violating) components of the decay amplitude.

The $F_j$ distributions describe the dynamics of the decay amplitudes and
are written as the product of an invariant mass term $R_j$, two Blatt--Weisskopf 
barrier form factors $X_J$, and an angular function $T_j$
\begin{equation}
F_j(m_{\rm max}^2,m_{\rm min}^2) \equiv R_j(m)\,X_J(p^{\star})\,X_J(q)\,T_j(m)\,,
\label{eq:Fj}
\end{equation}
where $m$ ($J$) is the mass (spin) of the resonance, $p^{\star}$ is the momentum
the bachelor pion that is not part of the resonance in the \B\ meson rest frame,
and $q$ is the momentum of either daughter in the rest frame of the resonance (we
use the $c=1$ convention for all equations in this paper). The $F_j$ are normalized
over the entire Dalitz plot:
\begin{equation}
\int\!\!\int \left|F_j(m_{\rm max}^2,m_{\rm min}^2)\right|^2 dm_{\rm max}^2 dm_{\rm min}^2 = 1\,.
\label{eq:FjNorm}
\end{equation}

The Blatt-Weisskopf barrier form factors~\cite{Blatt} are given by:
\begin{eqnarray}
X_{J=0}(z) & = & 1\,, \\
X_{J=1}(z) & = & \sqrt{1/[1 + (z \, r_{\rm{BW}})^2]}\,, \nonumber \\
X_{J=2}(z) & = & \sqrt{1/[(z \, r_{\rm{BW}})^4 + 3 (z \, r_{\rm{BW}})^2 + 9]}\,, \nonumber
\label{eq:BWFactors}
\end{eqnarray}
where the meson radius parameter $r_{\rm{BW}}$ is taken to be 
$4.0 \pm 1.0$ ({Ge\kern -0.1em V\!/}c)$^{-1}$~\cite{Amsler:2008zz}.

For most resonances in this analysis the $R_j$ are taken to be relativistic
Breit--Wigner lineshapes
\begin{equation}
R_j(m) = \frac{1}{(m^2_0 - m^2) - i m_0 \Gamma(m)}\,,
\label{eqn:BreitWigner}
\end{equation}
where $m_0$ is the nominal mass of the resonance and $\Gamma(m)$ is the mass-dependent width.
In the general case of a spin-$J$ resonance, the latter can be expressed as
\begin{equation}
\Gamma(m) = \Gamma_0 \left( \frac{q}{q_0}\right)^{2J+1} 
\frac{m_0}{m} \frac{X^2_J(q)}{X^2_J(q_0)}\,.
\label{eqn:resWidth}
\end{equation}
The symbol $\Gamma_0$ denotes the nominal width of the resonance.
The values of $m_0$ and $\Gamma_0$ are obtained
from standard tables~\cite{Amsler:2008zz} when they are well known.
The symbol $q_0$ denotes the value of $q$ when $m = m_0$.

The angular distribution terms $T_j$ in Eq.~(\ref{eq:Fj}) follow the Zemach tensor 
formalism~\cite{Zemach1,Zemach2}.
For the decay of a spin zero $B$-meson into a spin $J$ resonance and a spin zero bachelor particle 
this gives~\cite{Asner}
\begin{eqnarray}
T^{J=0}_j & = & 1\,, \\
T_j^{J=1} & = & -2 \, \vec{p}\cdot\vec{q}\,, \nonumber \\
T_j^{J=2} & = & \frac{4}{3} \left[3(\vec{p}\cdot\vec{q})^2 - (|\vec{p}\,||\vec{q}\,|)^2\right]\,, \nonumber
\label{eq:Zemach}
\end{eqnarray}
where $\vec{p}$ is the momentum of the bachelor particle and $\vec{q}$ is the
momentum of the resonance daughter with charge opposite from that of the
bachelor particle, both measured in the rest frame of the resonance.

The Gounaris--Sakurai parametrization~\cite{Gounaris:1968mw} of the $P$-wave
scattering amplitude for a broad resonance decaying to two pions is used for the
\rhoI\ and \rhoII\ lineshapes
\begin{equation}
R_{j}(m) = \frac{1 + \Gamma_0\,d/m_0}{(m_0^2 - m^2) + f(m) - i\, m_0 \Gamma(m)} \,
\label{eq:rhoGS}
\end{equation}
where
\begin{eqnarray}
\lefteqn{f(m) = \Gamma_0 \frac{m_0^2}{q_0^3} \times} \\
&& \left[ q^2 [h(m)-h(m_0)] + (m_0^2 - m^2) \, q^2_0 \, \frac{dh}{dm}\bigg|_{m_0} \right]\,, \nonumber
\label{eqn:fFun}
\end{eqnarray}
and the function $h(m)$ is defined as
\begin{equation}
h(m) = \frac{2}{\pi}\,\frac{q}{m}\,
       \ln\left(\frac{m+2q}{2m_\pi}\right)\,,
\label{eq:hFun}
\end{equation}
with
\begin{equation}
\frac{dh}{dm}\bigg|_{m_0} =
h(m_0)\left[(8q_0^2)^{-1}-(2m_0^2)^{-1}\right] \,+\, (2\pi m_0^2)^{-1}\,. 
\label{eq:dhdmFun}
\end{equation}
The normalization condition at $R_j(0)$ fixes the parameter
$d=f(0)/(\Gamma_0 m_0)$. It is found to be
\begin{equation}
d = \frac{3}{\pi}\frac{m_\pi^2}{q_0^2}\,
    \ln\left(\frac{m_0+2q_0}{2m_\pi}\right) 
    + \frac{m_0}{2\pi q_0} 
    - \frac{m_\pi^2 m_0}{\pi q_0^3}\,.
\label{eq:dFun}
\end{equation}

We model the nonresonant component using an empirical function that has been
found to accurately describe nonresonant contributions in other charmless three-body $B$
decays~\cite{Garmash:2004wa,Garmash:2005rv,Garmash:2006fh,Aubert:2006nu,Aubert:2007sd}:
\begin{eqnarray}
  A_{\rm nr} & = & c_{\rm nr} ( e^{-\alpha_{\rm nr} m_{\rm max}^2} + e^{-\alpha_{\rm nr} m_{\rm min}^2} ) \, .
\label{eq:nr}
\end{eqnarray}
We include this term in the coherent sum given by Eq.~(\ref{eq:totAmp}) when calculating 
the total signal amplitude over the Dalitz plot.

To allow comparison among experiments we present results also in terms of fit
fractions (${\it FF}_j$), defined as the integral of a single decay amplitude squared divided by the
total matrix element squared for the complete Dalitz plot
\begin{equation}
{\it FF}_j =
\frac
{\displaystyle\int\!\!\int{\left(\left|c_j F_j\right|^2 + \left|\cbar_j \Fbar_j\right|^2\right)} dm_{\rm max}^2\,dm_{\rm min}^2}
{\displaystyle\int\!\!\int{\left(\left|A\right|^2 + \left|\Abar\right|^2\right)} dm_{\rm max}^2\,dm_{\rm min}^2} \,.
\label{eq:fitfraction}
\end{equation}
Note that the sum of all the fit fractions is not necessarily unity due to the possible presence
of constructive or destructive interference. The $\CP$ asymmetry for each contributing resonance
is determined from the fitted parameters
\begin{eqnarray}
  \label{eq:cpasym}
  \CalACP_{\!,\,j} & = &
  \frac
  {\left|\cbar_j\right|^2 - \left|c_j\right|^2}
  {\left|\cbar_j\right|^2 + \left|c_j\right|^2} \\
  & = &
  \frac
  {-2\left(x_j \Delta x_j + y_j \Delta y_j\right)}
  {(x_j)^2 + (\Delta x_j)^2 + (y_j)^2 + (\Delta y_j)^2} \,\nonumber .
\end{eqnarray}

The signal Dalitz plot probability density function (PDF) is formed from
the total amplitude as follows:
\begin{eqnarray}
  \label{eq:SigDPLikeEqn}
  &&
  {\cal P}_{\rm sig}(m_{\rm max}^2,m_{\rm min}^2,q_{\B}) = \\ \nonumber
  &&
  \phantom{{\cal L}_{\rm sig}(m_{\rm max}^2}
  \frac
  {
    \frac{1+q_{\B}}{2} |A|^2 \; \varepsilon + \frac{1-q_{\B}}{2} |\Abar|^2 \; \epsilonbar
  }
  {
    \displaystyle{\int\!\!\int \!\! \left(\, |A|^2 \, \varepsilon +
    |\Abar|^2 \, \epsilonbar\,\right) dm_{\rm max}^2\,dm_{\rm min}^2}
  } \,,
\end{eqnarray}
where $q_{\B}$ is the charge of the \B-meson candidate, and
$\varepsilon \equiv \varepsilon(m_{\rm max}^2,m_{\rm min}^2)$ and 
$\epsilonbar \equiv \epsilonbar (m_{\rm max}^2,m_{\rm min}^2)$ are the signal reconstruction
efficiencies for \Bp\ and \Bm\ events, respectively, defined for all points in
the Dalitz plot. 

\section{Candidate Selection}

We reconstruct \B\ candidates from events that have four or more
charged tracks. Each track is required to be well measured and to
originate from the beam spot.
They must have a minimum
transverse momentum of 50\,\mevc, and a distance of closest approach
to the beam spot of less than 1.5\,\cm\ in the transverse plane and
less than 2.5\,\cm\ along the detector axis.
\B\ candidates are formed from combinations of three charged tracks,
and particle identification (PID) criteria are applied to reject
electrons and to separate pions from kaons.
In our final state, the average selection efficiency for pions that have
passed the tracking and PID requirements is about $93\%$ including geometrical
acceptance, while the average misidentification probability of kaons as
pions is close to $8\%$.

Two kinematic variables are used to identify signal \B\ decays.
The first variable is
\begin{equation}
\DeltaE = E_B^* - \sqrt{s}/2,
\end{equation}
the difference between the reconstructed CM energy of the \B-meson
candidate ($E_B^*$) and $\sqrt{s}/2$, where $\sqrt{s}$ is the total CM energy.
The second is the beam-energy-substituted mass
\begin{equation}
\mes = \sqrt{s/4 - |\vec{p}_B^*|^2},
\end{equation}
where $\vec{p}_B^*$ is the \B\ momentum measured in the CM frame.
The \mes\ distribution for signal events peaks near the \B\ mass with a
resolution of around $2.5\mevcc$, while the \DeltaE\ distribution peaks
at zero with a resolution of approximately $20\mev$.
We initially require events to lie in the region formed by the following
selection criteria: $5.200<\mes<5.286\gevcc$ and $-0.075<\DeltaE<0.300\gev$.
The region of \DeltaE\ below $-0.075\gev$ is heavily contaminated by four-body
\B\ decay backgrounds and is not useful for studying the continuum
background.
The selected region is then subdivided into three areas:
the ``left sideband'' ($5.20<\mes<5.26\gevcc$ and
$|\DeltaE|<0.075\gev$) used to study the background \DeltaE\ and
Dalitz plot distributions; 
the ``upper sideband'' ($5.230<\mes<5.286\gevcc$ and 
$0.1<\DeltaE<0.3\gev$) used to study the background \mes\ distributions;
and the ``signal region'' ($5.272<\mes<5.286\gevcc$ and
$|\DeltaE|<0.075\gev$) with which the final fit to data is performed.
Following the calculation of these kinematic variables, each of the
\B\ candidates is refitted with its mass constrained to the world-average
value of the \B\ meson mass~\cite{Amsler:2008zz} in order to improve the
Dalitz plot position resolution
and to make sure all events lie within the kinematic boundary of the
Dalitz plot.

The dominant source of background comes from light-quark and charm
continuum production ($\epem\to\qqbar$, where $q = u,d,s,c$). This
background is suppressed by requirements on event-shape variables
calculated in the CM frame.
We compute a neural network (NN) from the following five variables:
the ratio of the second- and zeroth-order angular moments ($\LtwoOverLzero$),
with $L_j=\sum_i p_i |\cos\theta_i|^j$, where $\theta_i$ is the angle of the 
track or neutral cluster $i$ with respect to the signal \B\ thrust axis,
$p_i$ is its momentum, and the sum excludes the daughters of the \B\ candidate;
the absolute value of the cosine of the angle between the 
direction of the \B\ and the detector axis; the magnitude of the cosine of the 
angle between the signal \B\ thrust axis and the detector axis; the output of a
multivariate \B-flavor tagging algorithm~\cite{Aubert:2004zt} multiplied by the
charge of the \B\ candidate; and the ratio of the measured proper
time difference of the two \B\ decay vertices and its statistical uncertainty.
We train the NN using samples of off-peak data and
signal Monte Carlo (MC) events generated with the phase-space distribution.
A selection requirement is imposed on the NN output that accepts about $48\%$
of signal events while rejecting $97\%$ of continuum background events.

Dalitz plot distributions of the reconstruction efficiency for \Bp\ and \Bm
events are modeled with two-dimensional histograms formed from a sample of
around $7\times10^{6}$ \BtoPPP\ phase-space MC events.
All selection criteria are applied except for the exclusion of certain
invariant-mass regions described below.
We take the ratio of two histograms, the denominator containing the
true Dalitz plot distribution of all generated MC events and the numerator
containing the reconstructed MC events. The reconstructed events are
weighted in order to correct for differences between data and MC simulations in the
tracking and PID efficiencies.
In order to give better resolution near the edges of the Dalitz plot, where
most reconstructed events lie, the histograms are formed in the ``square
Dalitz plot''~\cite{Aubert:2007jn,Aubert:2005sk} coordinates.
We use 50$\times$50 bins and smooth these histograms by applying linear
interpolation between neighboring bins.
The efficiency shows very little variation across most of the
Dalitz plot
but decreases towards the corners where one of the particles has low
momentum.
The effect of experimental resolution on the signal model is neglected
since the resonances under consideration are sufficiently broad.
The average reconstruction efficiency for events in the signal region for the
phase-space MC sample is about $15\%$.
The fraction of misreconstructed signal events is only $5\%$,
and MC studies indicate that there is no need for any explicit treatment of
these events.

\section{Backgrounds}
\label{sec:backgrounds}

In addition to the continuum (\qqbar) background we also have backgrounds
from \BB\ events. There are four main sources:
(i) combinatorial background from three unrelated tracks;
(ii) three- and four-body \B\ decays involving an intermediate $D$ meson;
(iii) charmless two- and four-body decays with an extra or missing particle; and
(iv) three-body decays with one or more particles misidentified.
We reject background from two-body decays of $D$ mesons and charmonium 
states by excluding invariant masses (in units of
$\rm{Ge\kern -0.1em V\!/}c^2$) in the ranges:
$1.660 < m_{\pipi} < 1.920$,
$3.051 < m_{\pipi} < 3.222$, and
$3.660 < m_{\pipi} < 3.820$.
These ranges reject decays from \DzbtoKpi\ (or \pipi), \JPsitoll, and \Psitoll\
respectively, where $\ell$ is a lepton that has been misidentified as a pion.
We also employ a special requirement to reject the decay process
$\B^{\pm}\to\KS\pi^{\pm};\KS\to\pip\pim$, by excluding candidates where
the vertexed mass of two oppositely charged pions lies in the range of
$[478,516]\mevcc$.

We use a large sample of MC-simulated \BB\ decays, equivalent to approximately
$3$ times the integrated luminosity of the data sample, to identify the
important \B\ backgrounds that survive the invariant-mass exclusion requirements
described above. In total, $53$ \B-meson decay modes are identified for which
larger samples of exclusive MC events are used for further study. We combine
modes that have similar behavior in the discriminating variables \mes\ and
\DeltaE\ into a \B-background category. There are four such categories: the
first contains the two-body decays \Bztopippim\ and \BztoKmpip, the second
is dominated by \BtoKPP\ and contains other decays with similar topologies,
the third contains only \BtoPPPz, and the fourth contains the remaining
backgrounds from \B\ decays that are combinatorial in nature. For each
\B-background category the combined \mes, \DeltaE, and Dalitz plot distributions
are created where the relative contributions of various decay modes in a
specific category are calculated from the reconstruction efficiencies from MC
simulations and the branching fractions listed by the Particle Data
Group~\cite{Amsler:2008zz} and the Heavy Flavor Averaging Group~\cite{Barberio:2008fa}.
These distributions are used in the likelihood fit described below.

Background Dalitz plot distributions are included in the likelihood fit
through the use of two-dimensional histograms. For backgrounds from \B
decays these histograms are formed from the various MC samples.
For the continuum background the left sideband data sample is used.
Since this data sideband also contains events from \B\ decays, MC samples
are used to subtract these events. To these \B-subtracted
sideband events, we add off-peak data events from across the whole range
of \mes\ and \DeltaE\ in order to enhance statistics.
We have verified that the shapes of various discriminating
variables are compatible between the sideband and off-peak events.
As for the reconstruction efficiency histograms, the background Dalitz plot
distributions are formed in the square Dalitz plot coordinates and are
smoothed by linear interpolation applied between neighboring bins.
Separate histograms are constructed for \Bp\ and \Bm\ events.
The \qqbar- and \B-background PDFs are identical in their construction, and
the \qqbar\ PDF is shown here as an example:
\begin{eqnarray}
\label{eq:BgDPLikeEqn}
&& \hspace{-7mm} 
{\cal P}_{\qqbar}(m_{\rm max}^2,m_{\rm min}^2,q_{\B}) = 
\frac{1}{2} (1 - q_{\B} {\cal A}_{\qqbar}) \times \\ \nonumber
&& \left( 
  \frac{
    \frac{1+q_{\B}}{2} \; Q(m_{\rm max}^2,m_{\rm min}^2)
  }{
    \displaystyle{\int\!\!\int 
      Q(m_{\rm max}^2,m_{\rm min}^2) \; dm_{\rm max}^2 \, dm_{\rm min}^2} 
  } + 
\right. \\ \nonumber
&& \phantom{\Bigg(++++} \left. 
  \frac{
    \frac{1-q_{\B}}{2} \; \Qbar(m_{\rm max}^2,m_{\rm min}^2)
  }{
    \displaystyle{\int\!\!\int 
      \Qbar(m_{\rm max}^2,m_{\rm min}^2) \; dm_{\rm max}^2 \, dm_{\rm
        min}^2}
  }
\right)\,,
\end{eqnarray}
where ${\cal A}_{\qqbar}$ is the charge asymmetry in the
background, and $Q(m_{\rm max}^2,m_{\rm min}^2)$ and $\Qbar(m_{\rm
 max}^2,m_{\rm min}^2)$ are the Dalitz plot distributions of \qqbar
events in selected \Bp\ and \Bm\ samples, respectively.

\section{Maximum Likelihood Fit}

To provide further discrimination between the signal and background
hypotheses in the likelihood fit, we include PDFs for the kinematic
variables \mes\ and \DeltaE, which multiply that of the Dalitz plot.
The signal \mes\ shape is modeled with the sum of a Gaussian function 
and a Crystal-Ball lineshape~\cite{crystalBall},
and the \DeltaE\ shape is modeled with a double Gaussian function. The
parameters of these functions are obtained from a sample of \BtoPPP\ MC
events, modeled according to the Dalitz plot distribution from
Ref.~\cite{Aubert:2005sk}, and are appropriately adjusted to account for
possible differences between data and MC simulations determined with a control
sample of \BptoDzbpip; \DzbtoKpi\ decays. These parameters are fixed in the fit
to data.

The \qqbar\ \mes\ distribution is modeled with the experimentally motivated
ARGUS function~\cite{Albrecht:1990am}. 
The end point for the ARGUS function is fixed to
$5.289\gevcc$, and the parameter describing the shape is fixed to the value
determined from the combined sample of upper sideband and off-peak data.
We model the continuum \DeltaE\ shape using a linear function, the slope
of which is fixed to the value determined from the left sideband and off-peak data.
The \BB\ background distributions are modeled with histograms obtained
from the mixture of \BB\ MC samples.
The yields of signal and \qqbar\ events are allowed to vary in the final fit
to the data while the yields of \BB\ backgrounds are fixed to
11 (two-body decays), 195 (\BtoKPP\ type), 117 (\BtoPPPz), and 495 (combinatorial)
events.

The complete likelihood function is given by:
\begin{eqnarray}
{\cal L} &=& e^{-N} \times \\ \nonumber
&&
\prod_j^{N_e}
\Bigg[
\sum_k N_k {\cal P}_k^j(m_{\rm max}^2,m_{\rm min}^2,\mes,\DeltaE,q_{\B})
\Bigg] \,,
\end{eqnarray}
where $N$ is equal to $\sum_k{N_k}$, $N_k$ is the yield for the event category $k$, 
$N_e$ is the total number of events in the data sample, 
and ${\cal P}_k^j$ is the PDF for the category $k$
for event $j$, which consists of a product of the Dalitz plot, \mes, and \DeltaE\ PDFs.
The function $-2\ln{\cal L}$ is minimized in an unbinned fit to the data.

Our nominal signal Dalitz plot model comprises a momentum-dependent
nonresonant component and four intermediate resonance states:
\rhoIpipm, \rhoIIpipm, \fIIpipm, and \fIIIpipm. The parameters used to describe these 
states are summarized in \tabref{params}. 
We fit $4335$ $B$ candidates in the signal region selected from the data 
to obtain the central values of the $x_j$,
$\Delta x_j$, $y_j$, and $\Delta y_j$ parameters for each component, and use
Eqs.~(\ref{eq:fitfraction}) and~(\ref{eq:cpasym}) to calculate the fit
fractions and \CP\ asymmetries. 
We use \rhoIpipm\ as the reference amplitude, fixing its $x$, $y$, 
and $\Delta y$ parameters to unity, zero, and zero, respectively.
The signal yield, \qqbar\ background yield and asymmetry are also free
parameters of the fit, giving a total of 20 free parameters.

The Dalitz plot model was determined using the results
of our previous analysis~\cite{Aubert:2005sk} and the changes
in the fit likelihood and $\chi^2$ values when omitting or adding resonances.
The latter is calculated from the projection of the fit results onto the
Dalitz plot using the formula
\begin{equation}
\chi^2 = \sum_{i=1}^{n_b} \frac{[y_i - f(x_i)]^2}{f(x_i)},
\label{eq:chisq}
\end{equation}
where $y_i$ is the number of data events in bin $i$ and $f(x_i)$ is the
number of events in that bin as predicted by the fit result. 
The number of degrees of freedom is calculated as $n_b - h - 1$, 
where $n_b$ is the total number of bins used and $h$ is the number of free
parameters in the fit. A minimum of 20 entries in each bin is required; if
this requirement is not met then the neighboring bins are combined.  Typically,
$n_b$ takes values around 100. 

In our previous study we found significant contributions from \rhoIpipm\ and
\fIIpipm; with \fIpipm, \rhoIIpipm, and a uniform nonresonant term also
included in the model.
Because of the larger data sample and many improvements to the analysis,
we find it necessary to include an additional contribution from \fIIIpipm, and
to use a momentum-dependent nonresonant amplitude [see \eqref{nr}] in order to
achieve a 
reasonable agreement of the fit with the data.
We do not find any significant signal from \fIpipm, so we exclude this channel
from our nominal model and calculate an upper limit for its fit fraction.
The statistical significance of the presence of a component is estimated by
evaluating the difference $\Delta\ln{\cal L}$ between the negative
log-likelihood of the nominal fit and that of a fit where all of the $x$, $y$,
$\Delta x$, and $\Delta y$ parameters for the given component are fixed to
zero. This is then used to evaluate a $p$ value
\begin{equation}
p = \int_{2\Delta\ln{\cal L}}^{\infty} f(z;n_d) \,dz \,,
\end{equation}
where $f(z;n_d)$ is the PDF of the \chisq\ distribution and $n_d$ is the number of degrees of
freedom, four in this case.
We then determine the equivalent one-dimensional significance from this
$p$ value.
We find that the \fII\ contribution has a statistical significance of
$6.1\sigma$, the \rhoII\ $4.6\sigma$ and the \fIII\ $3.9\sigma$.

Since the mass and width of the \fIII\ state are not well
known~\cite{Amsler:2008zz}, we determine the preferred values from data by
scanning the likelihood values obtained with different parameters.
The mass and width are determined to be
$m_{f_0(1370)} = 1400\pm40\mevcc$ and $\Gamma_{f_0(1370)} = 300\pm80\mev$,
with a correlation of $(-39\pm4)\%$, where the errors are statistical only,
and are obtained from a fit to the two-dimensional likelihood profile.
Similarly, we determine the parameter of the nonresonant lineshape to be 
$\alpha_{\rm nr} = 0.28\pm0.06\gevccSqinv$ (statistical uncertainties only).

Possible contributions from $\chiczero\pipm$ and $\chictwo\pipm$
are not significant so we set upper limits on their branching
fractions. Furthermore, we do not find any evidence for a very broad
enhancement at low $\pipi$  invariant mass such as could be caused by the 
decay $\Bpm\to\sigma\pipm$.

\begin{table}[hbt!]
 \caption{
  Parameters used to describe intermediate states in our nominal model.
  GS and RBW refer to the Gounaris-Sakurai and relativistic Breit-Wigner
  lineshapes, respectively.
 }
 \label{tab:params}
 \begin{tabular}{ccccc}
  \hline
  \hline
  Resonance & Lineshape & Mass (MeV\!/$c^2$) & Width (MeV) & Ref. \\
  \hline
  \rhoI     &   GS      & $775.49\pm0.34$ & $149.4\pm1.0$ & \cite{Amsler:2008zz} \\
  \rhoII    &   GS      & $1465\pm25$ & $400\pm60$ & \cite{Amsler:2008zz} \\
  \fII      &   RBW     & $1275.1\pm1.2$ & $185.0_{-2.4}^{+2.9}$ & \cite{Amsler:2008zz} \\
  \fIII     &   RBW     & $1400\pm40$ & $300\pm80$ & \hspace{-2mm}See text \\  
  \hline
  \hline
 \end{tabular}
\end{table}

\begin{figure}[!htb]
  \begin{center}
    \includegraphics[width=0.494\columnwidth]{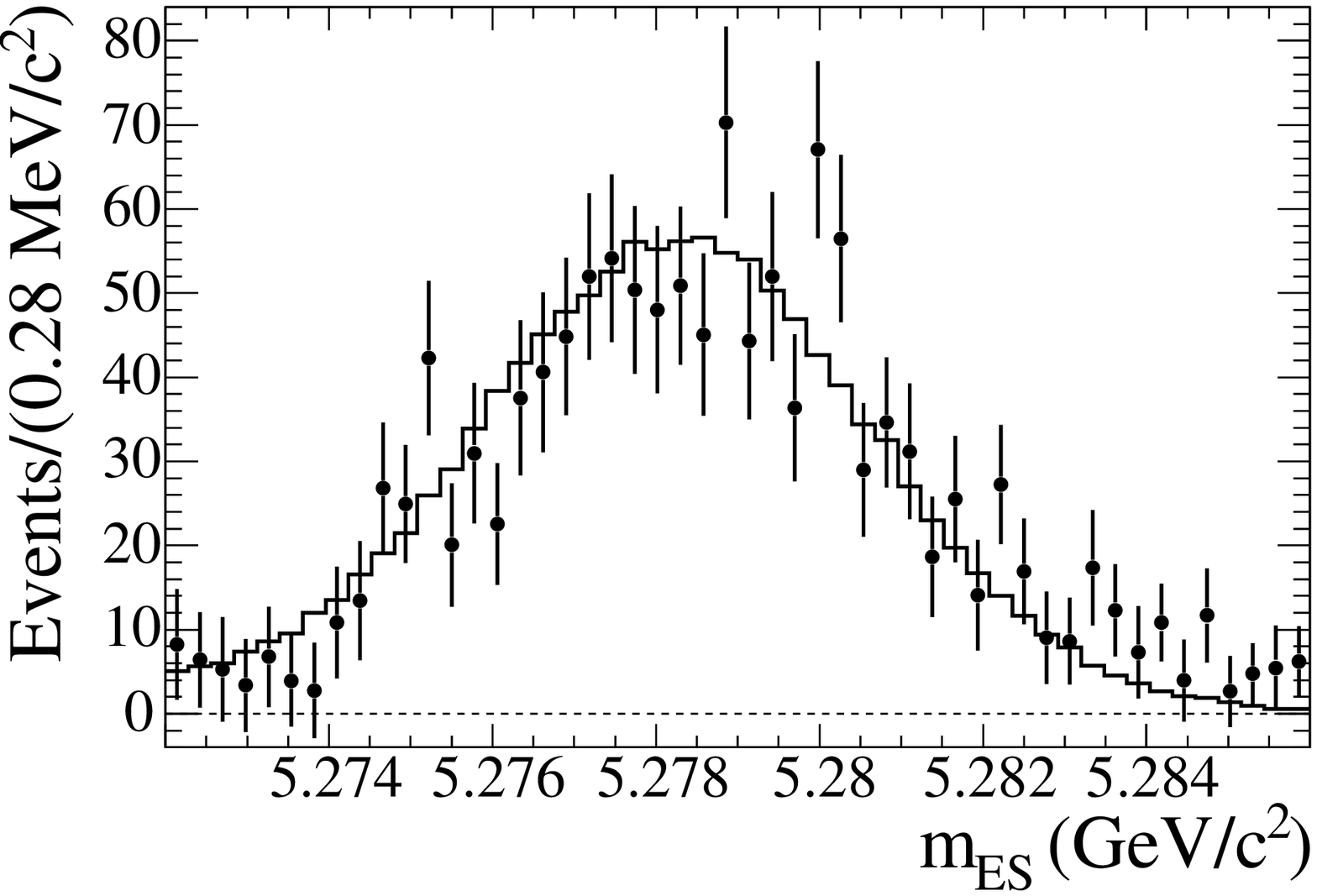}
    \includegraphics[width=0.494\columnwidth]{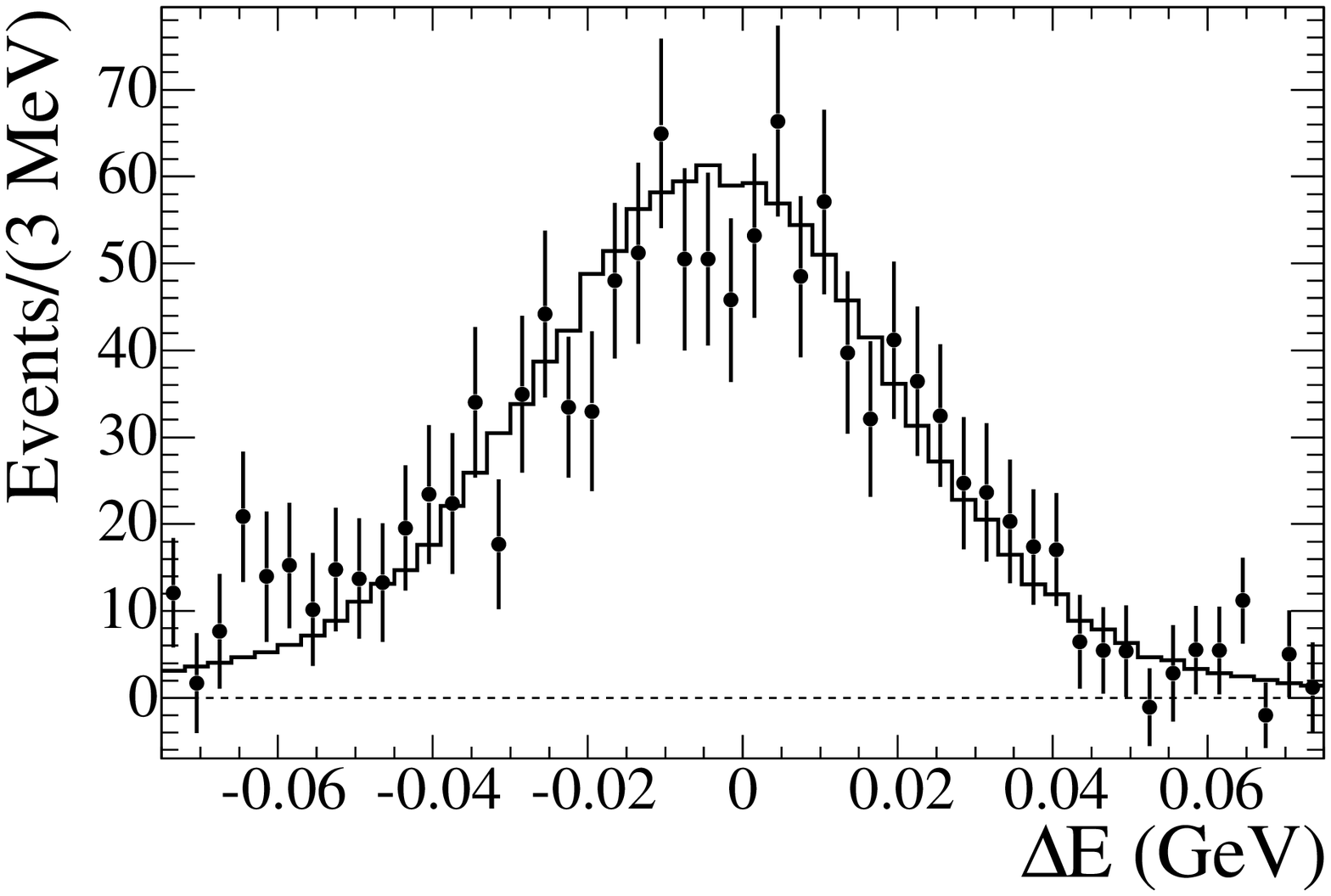}
    \includegraphics[width=0.494\columnwidth]{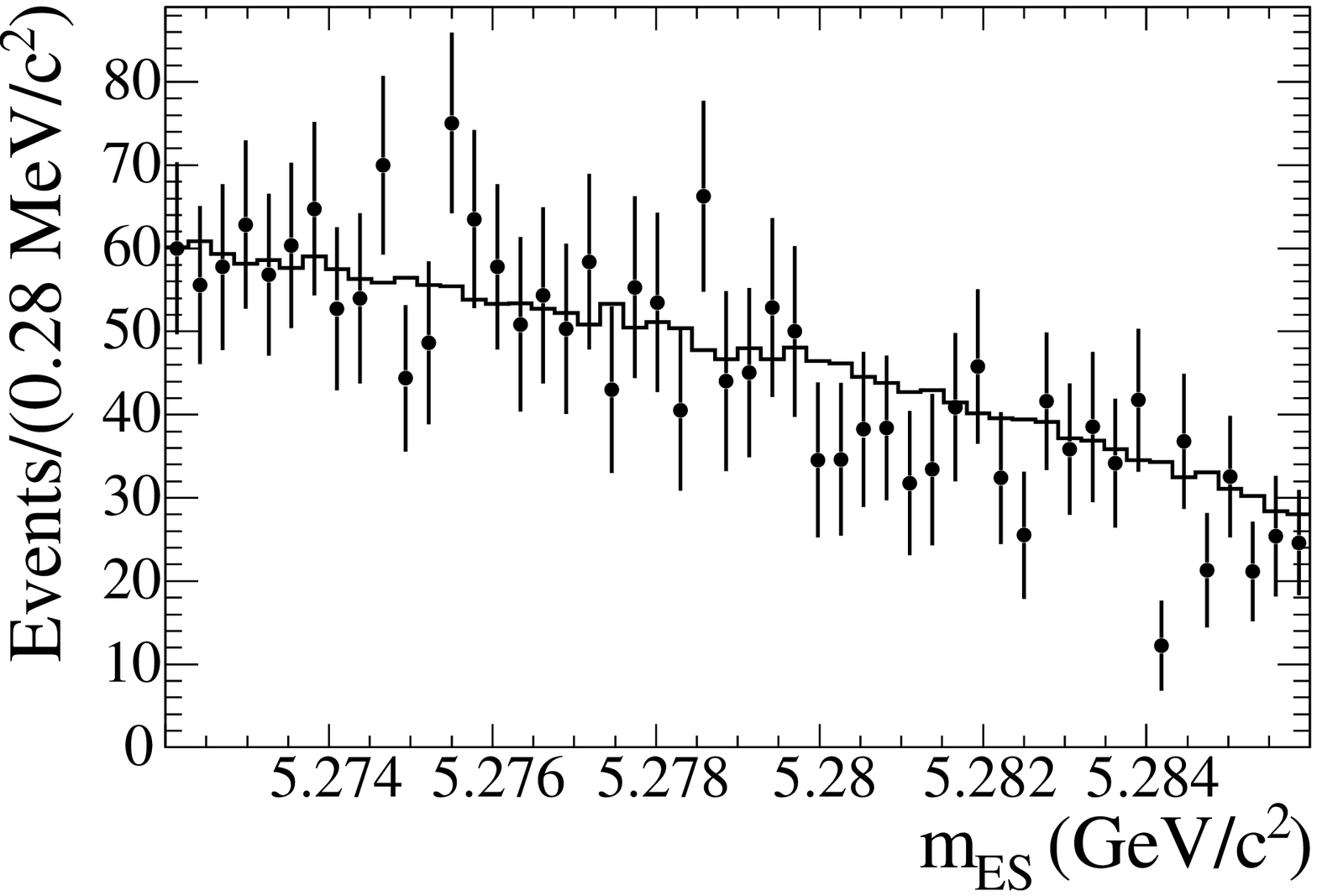}
    \includegraphics[width=0.494\columnwidth]{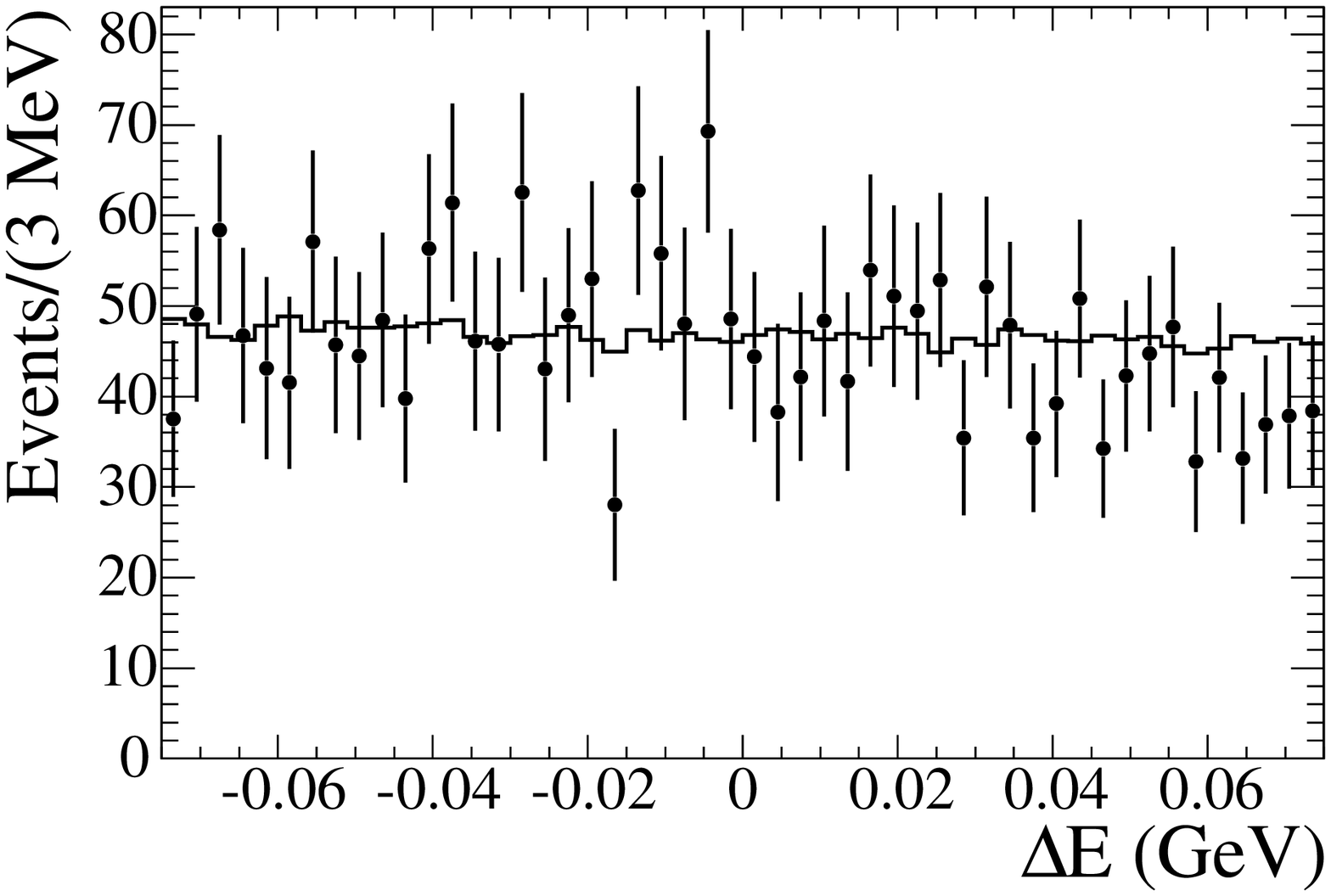}
    \caption{
      (Top) signal and (bottom) \qqbar\ distributions of (left) \mes\ and
      (right) \DeltaE\ obtained from the fit to data using event-by-event
      signal and \qqbar\ background probabilities~\cite{Pivk:2004ty}.
      The solid lines show the PDF shapes used in the fit.
    }
    \label{fig:mesde-splots}
  \end{center}
\end{figure}

\begin{figure}[!htb]
  \begin{center}
    \includegraphics[width=0.95\columnwidth]{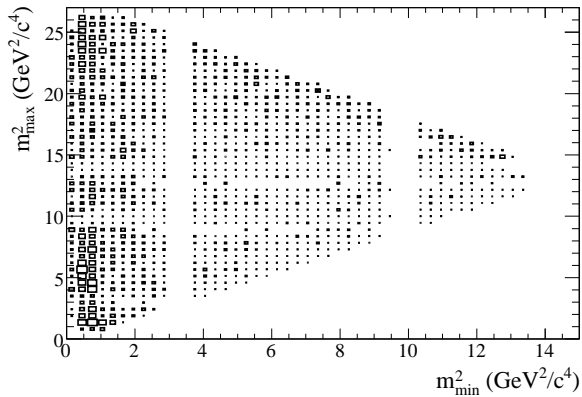}
    \caption{
      Background-subtracted Dalitz plot of the combined \BtoPPP\ data sample
      in the signal region. The plot shows bins with greater than zero
      entries. The area of the boxes is proportional to the number of
      entries. The depleted bands are the charm and charmonia exclusion regions.
    }
    \label{fig:dp}
  \end{center}
\end{figure}

Figure~\ref{fig:mesde-splots} shows the \mes\ and \DeltaE\ distributions of
signal and \qqbar\ background determined from the fit with event-by-event
signal and \qqbar\ background probabilities for each candidate event~\cite{Pivk:2004ty}.
The background-subtracted Dalitz plot of the data in the signal region
can be seen in \shortfigref{dp}. 
The $\chi^2$ per number of degrees of freedom of the projection of the fit
result onto the Dalitz plot is 82/84.
Using the fitted signal distribution we calculate the average reconstruction
efficiency for our signal sample to be $18\%$.

We generate a large number of MC experiments with the fitted parameters, and
from the spread of results of fits to those experiments we determine the
statistical uncertainties on the parameters, ${\it FF}_j$,
and $\CalACP_{\!,\,j}$. 
This procedure takes into account correlations between the 
$x_j$, $\Delta x_j$, $y_j$, and $\Delta y_j$ parameters.
The linear correlation coefficients between the ${\it FF}_j$ and
$\CalACP_{\!,\,j}$ parameters are also obtained and are presented in
\appref{correlations}.
In order to calculate the branching fraction for an intermediate mode,
we multiply the fit fraction of the latter by the total inclusive
$\Bpm\to\pipm\pipm\pimp$ branching fraction. They are needed for
comparison with previous measurements and theoretical predictions.
The \PPP\ signal yield is found to be $1219\pm 50\pm 75\,^{+29}_{-24}$ events and the
inclusive \CP\ asymmetry to be $\left(+3.2\pm 4.4\pm 3.1\,^{+2.5}_{-2.0}\right)\%$, where the
uncertainties are statistical, systematic, and model-dependent, respectively.
Additionally, the total yield and \CP\ asymmetry of the continuum background are found to be
$2337\pm 62$ events and $(+0.2\pm 2.7)\%$, respectively, where the uncertainties are 
statistical only. Further results are shown in Tables~\ref{tab:summarytable}
and~\ref{tab:conclusion-results-summary}.

\begin{figure}[!htb]
  \begin{center}
    \includegraphics[width=0.494\columnwidth]{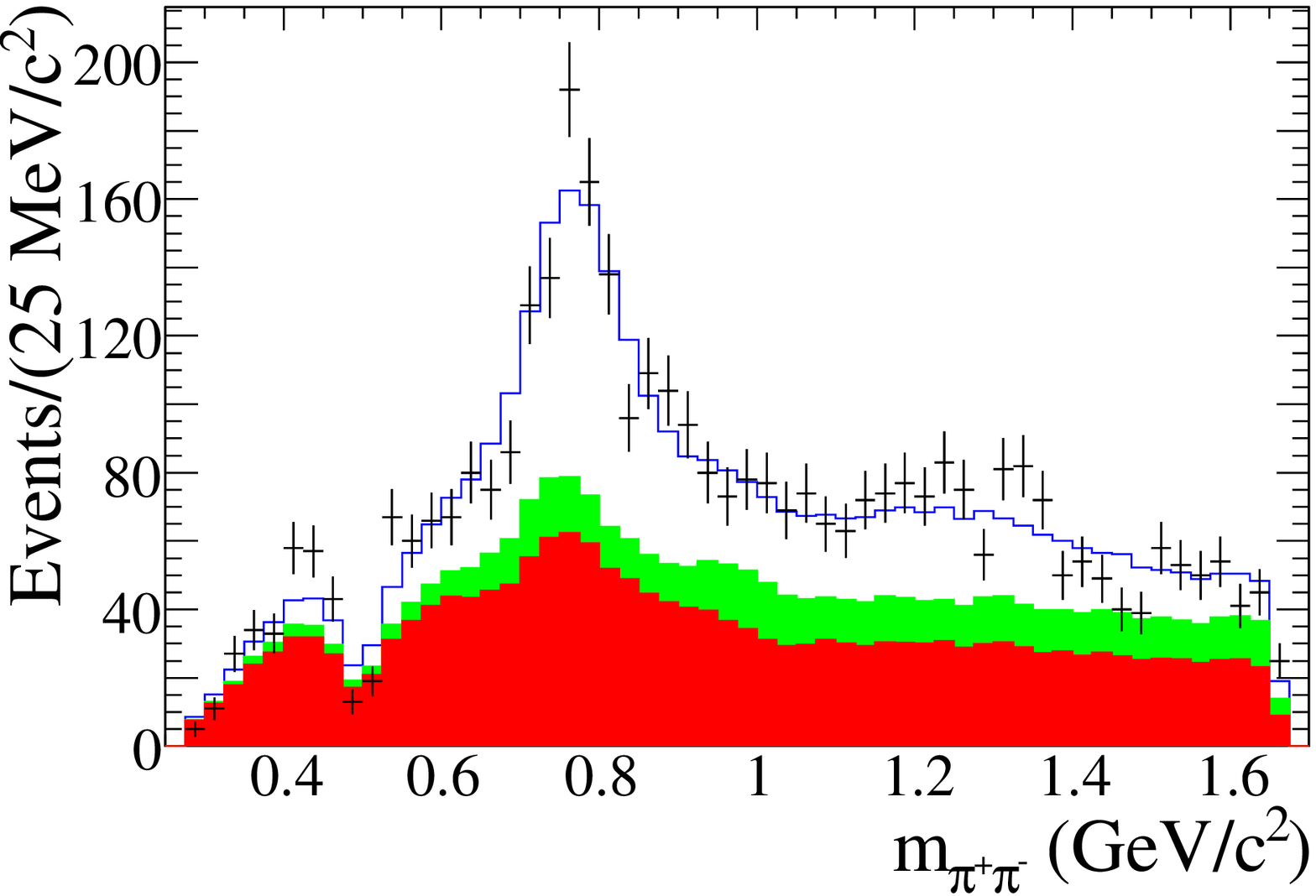}
    \includegraphics[width=0.494\columnwidth]{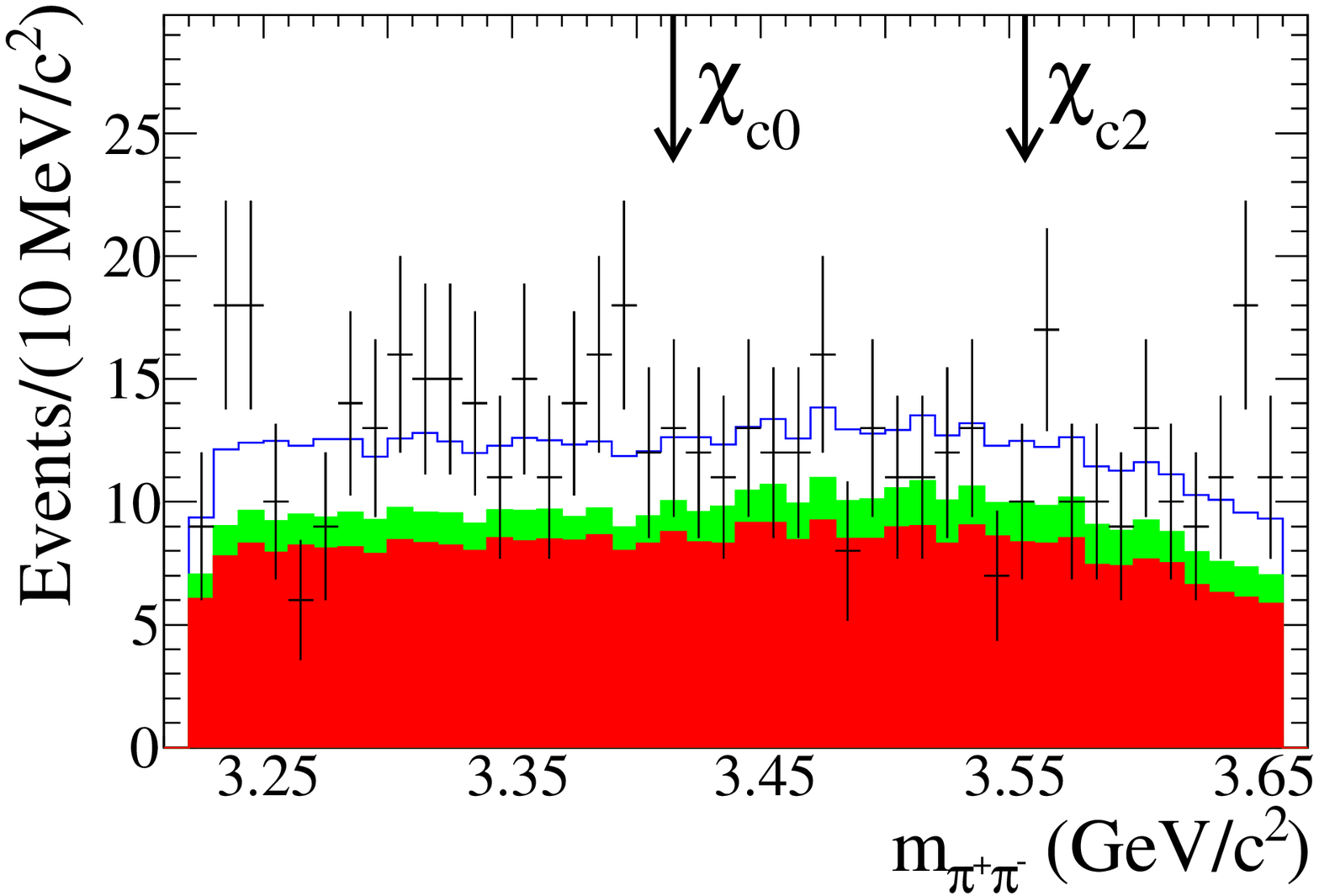}
    \caption{(color online)
      Dipion invariant mass projections: 
      (left) in the \rhoI\ region; and
      (right) in the regions of \chiczero\ and \chictwo.
      The data are the points with statistical error bars,
      the dark-shaded (red) histogram is the \qqbar\ component, 
      the light-shaded (green) histogram is the \BB\ background
      contribution, while the upper (blue) histogram shows the total fit result. 
      The dip near 0.5\gevcc in the left plot is due to the rejection of
      events containing \KS\ candidates.
    }
    \label{fig:nom}
  \end{center}
\end{figure}

\begin{figure}[!htb]
  \begin{center}
    \includegraphics[width=0.98\columnwidth]{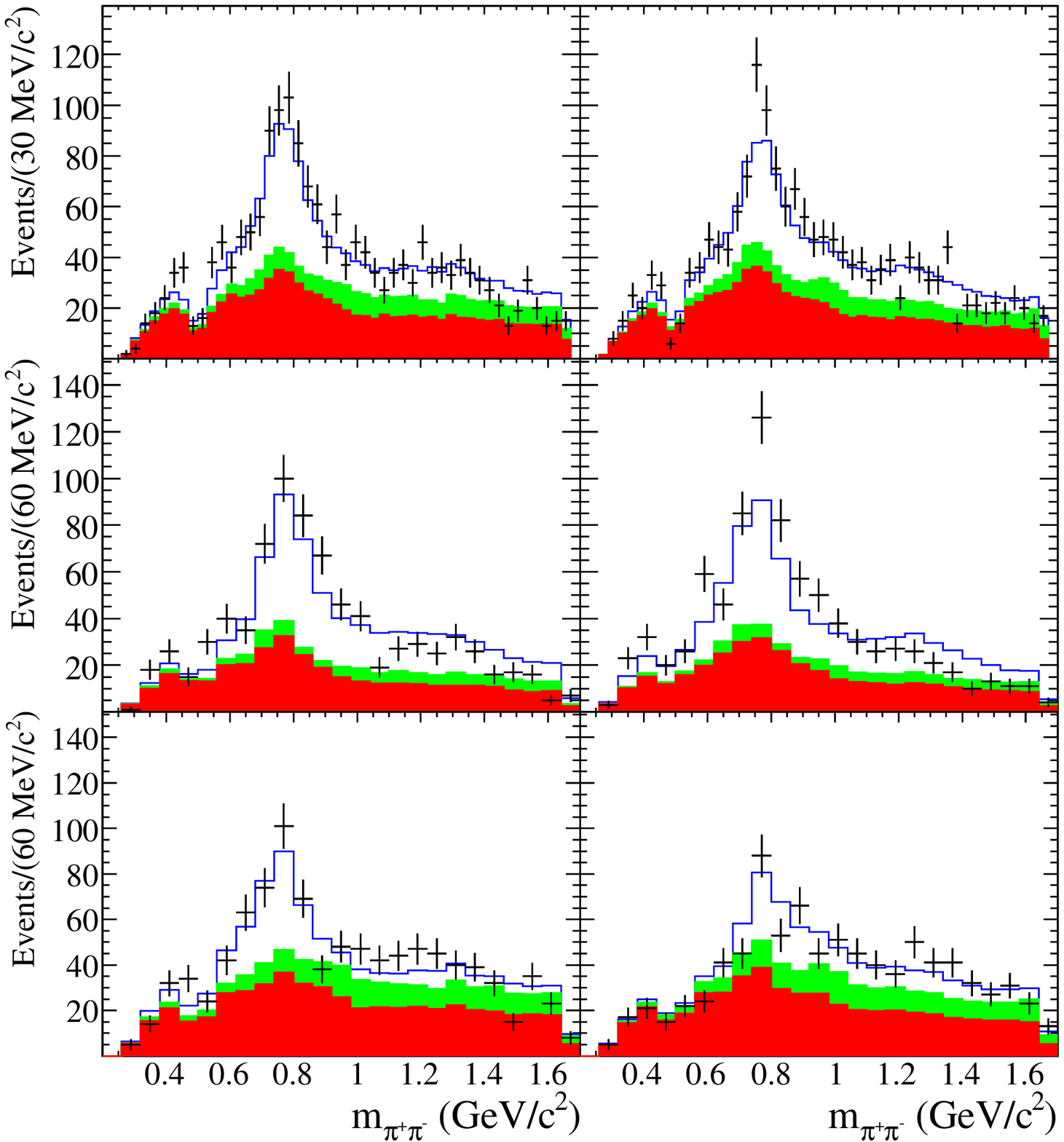}
    \caption{(color online)
      Dipion invariant mass projection in the \rhoI\ region for 
      (left) $B^-$ and (right) $B^+$ candidates. 
      The top row shows all candidates, 
      the middle row shows those with $\cos\theta_{H}>0$ and 
      the bottom row shows those with $\cos\theta_{H}<0$.
      The colors and shadings follow the same convention as \shortfigref{nom}.
    }
    \label{fig:nom-cpv}
  \end{center}
\end{figure}

Projections of the data, with the fit result overlaid, as $\pipm\pimp$
invariant-mass distributions can be seen in \shortfigref{nom}.
A detailed examination of possible direct \CP\ violation effects in the low
$\pipm\pimp$ invariant-mass region is shown in \shortfigref{nom-cpv},
where we have subdivided the data into positive and negative values of
$\cos\theta_H = \vec{p}\cdot\vec{q}\,/\!\left(|\vec{p}\,||\vec{q}\,|\right)$,
where $\theta_H$ is the helicity angle, $\vec{p}$ is the momentum of the bachelor particle 
and $\vec{q}$ is the momentum of the resonance daughter with charge opposite from that of the
bachelor particle, both measured in the rest frame of the resonance.
The agreement between the fit result and the data is generally good; the
$\chi^2$ per number of non-zero bins for these plots varies between 35/46 and
34/24. 

We calculate $90\%$ confidence-level (CL) upper limits for components not
included in the nominal Dalitz plot model. These are obtained by generating
many MC experiments from the results of fits to the data where the extra
component is added to the nominal Dalitz plot model, with all major systematic
sources varied within their $1\sigma$ uncertainties. 
We fit these MC samples and plot the fit fraction distributions.
The $90\%$ CL upper limit for each fit fraction is the value which includes $90\%$
of the MC experiments. The branching fraction upper limit is then the product of
the fit fraction upper limit and the total branching fraction for \BtoPPP.

\renewcommand{\arraystretch}{1.5}

\begin{table*}
  \caption{Results of fits to data, with statistical, systematic and model-dependent uncertainties.}
  \label{tab:summarytable}
  \resizebox{\textwidth}{!}{
    \begin{tabular}{l r@{$\pm$}c@{$\pm$}c@{$\,$}l r@{$\pm$}c@{$\pm$}c@{$\,$}l r@{$\pm$}c@{$\pm$}c@{$\,$}l r@{$\pm$}c@{$\pm$}c@{$\,$}l}
      \hline
      \hline
Resonance    & \fcc{$x$}                                     & \fcc{$y$}                                     & \fcc{$\Dx$}                                   & \fcc{$\Dy$}                                   \\
\hline
\vspace{1mm}
$\rhoI\pipm$  & \fcc{1.0 (fixed)}                             & \fcc{0.0 (fixed)}                             & $-0.092$&$0.036$&$0.027$&$^{+0.071}_{-0.012}$ & \fcc{0.0 (fixed)}                             \\
\vspace{1mm}
$\rhoII\pipm$ & $-0.292$&$0.071$&$0.065$&$^{+0.182}_{-0.054}$ & $ 0.175$&$0.078$&$0.048$&$^{+0.133}_{-0.042}$ & $ 0.109$&$0.080$&$0.059$&$^{+0.038}_{-0.116}$ & $ 0.211$&$0.073$&$0.038$&$^{+0.032}_{-0.146}$ \\
\vspace{1mm}
$\fII\pipm$   & $ 0.136$&$0.064$&$0.040$&$^{+0.178}_{-0.029}$ & $ 0.149$&$0.052$&$0.030$&$^{+0.022}_{-0.077}$ & $ 0.101$&$0.063$&$0.016$&$^{+0.031}_{-0.183}$ & $-0.248$&$0.052$&$0.024$&$^{+0.024}_{-0.026}$ \\
\vspace{1mm}
$\fIII\pipm$  & $ 0.397$&$0.067$&$0.058$&$^{+0.047}_{-0.050}$ & $-0.151$&$0.081$&$0.052$&$^{+0.057}_{-0.187}$ & $-0.387$&$0.064$&$0.029$&$^{+0.072}_{-0.082}$ & $-0.168$&$0.086$&$0.055$&$^{+0.160}_{-0.046}$ \\
\vspace{1mm}
Nonresonant  & $-0.200$&$0.091$&$0.029$&$^{+0.239}_{-0.045}$ & $-0.682$&$0.070$&$0.038$&$^{+0.032}_{-0.082}$ & $-0.392$&$0.089$&$0.055$&$^{+0.037}_{-0.128}$ & $ 0.046$&$0.069$&$0.055$&$^{+0.101}_{-0.124}$ \\
\hline
\hline
\end{tabular}
}
\end{table*}

\renewcommand{\arraystretch}{1}

\renewcommand{\arraystretch}{1.5}

\begin{table*}[hbt]
\caption
{Summary of measurements of branching fractions (averaged over charge
conjugate states) and \CP\ asymmetries.
The first error is statistical, the second is systematic and the third
represents the model dependence.
Also included are $90\%$ CL upper limits
of the branching fractions of the components that do not have
statistically significant fit fractions.
}
\label{tab:conclusion-results-summary}
\begin{tabular}{lccc}
\hline
\hline
Mode                           & Fit Fraction (\%)                  & $\BR(\Bpm \to {\rm Mode}) (10^{-6})$ & $\CalACP$ (\%)                    \\
\PPP\ Total                    &                                    & $15.2\pm0.6\pm1.2\,^{+0.4}_{-0.3}$  & $+3.2\pm4.4\pm3.1\,^{+2.5}_{-2.0}$ \\
\hline
\rhoIpipm; \rhoItopippim       & $53.2\pm3.7\pm2.5\,^{+1.5}_{-7.4}$ & $8.1\pm0.7\pm1.2\,^{+0.4}_{-1.1}$   & $+18\pm7\pm5\,^{+2}_{-14}$         \\
\rhoIIpipm; \rhoIItopippim     & $9.1\pm2.3\pm2.4\,^{+1.9}_{-4.5}$  & $1.4\pm0.4\pm0.4\,^{+0.3}_{-0.7}$   & $-6\pm28\pm20\,^{+12}_{-35}$       \\
\fIIpipm; \fIItopippim         & $5.9\pm1.6\pm0.4\,^{+2.0}_{-0.7}$  & $0.9\pm0.2\pm0.1\,^{+0.3}_{-0.1}$   & $+41\pm25\pm13\,^{+12}_{-8}$       \\
\fIIIpipm; \fIIItopippim       & $18.9\pm3.3\pm2.6\,^{+4.3}_{-3.5}$ & $2.9\pm0.5\pm0.5\,^{+0.7}_{-0.5}\,\,(<4.0)$ & $+72\pm15\pm14\,^{+7}_{-8}$\\
\PPP\ nonresonant              & $34.9\pm4.2\pm2.9\,^{+7.5}_{-3.4}$ & $5.3\pm0.7\pm0.6\,^{+1.1}_{-0.5}$   & $-14\pm14\pm7\,^{+17}_{-3}$        \\
\fIpipm; \fItopippim           & -                                  & $<1.5$                              & -                                  \\
\chiczpipm; \chicztopippim     & -                                  & $<0.1$                              & -                                  \\
\chictwopipm; \chictwotopippim & -                                  & $<0.1$                              & -                                  \\
\hline
\hline
\end{tabular}
\end{table*}

\renewcommand{\arraystretch}{1}

We have searched for the presence of multiple solutions in the fit to data
with the nominal model. We find a second solution with a value of 
$-2\ln{\cal L}$ about ten units higher than our nominal fit, 
and with a $\chi^2$ of the Dalitz plot projection increased by four units.
A comparison of the results between the two solutions is given in \appref{secondSoln}.
The most significant difference is seen in the \fIII\ fit fraction,
which is much smaller in the second solution.
In Dalitz plot analyses of $K\pi\pi$ and $KKK$ final states~\cite{Aubert:2005ce,Aubert:2008bj,Aubert:2007vi,Garmash:2004wa,Garmash:2005rv,Garmash:2006fh,Aubert:2006nu,Aubert:2007sd},
similar phenomena relating to multiple solutions have been observed,
and interpreted as being due to differences in the possible interference
pattern -- constructive or destructive interference --
between nonresonant and broad amplitudes.
Interference between the nonresonant and \fIII\ amplitudes
appears to be a plausible explanation for the effect in this analysis.
As can be seen in \shortfigref{nom-cpv}, the data exhibit a non-trivial
interference pattern, and the difficulty in modeling this effect
leads to model uncertainties in our results.

\section{Systematic Uncertainties}

Systematic uncertainties that affect the measurement of fit
fractions, phases, event yields, and \CP\ asymmetries are summarized in Table~\ref{tab:systematics-table}.
The fixed \BB-background yields and asymmetries 
are allowed to vary and the variations of the other fitted parameters are
taken as the uncertainties.
The effect of limited statistics of the data sideband and MC samples
used to obtain the fixed shapes of all the histogram PDFs is accounted
for by fluctuating independently the histogram bin contents in accordance
with their errors and repeating the nominal fit.
The uncertainties on how well the samples model these distributions
are also taken into account through various cross-checks, including
variation of the mass rejection ranges and comparison of continuum shapes
between sideband and signal region in MC samples.

\renewcommand{\arraystretch}{1.5}

\begin{table*}[htb]
\caption
{Absolute maximum values of the systematic uncertainties for the amplitude coefficients, 
fit fractions, signal yield, and \CP\ asymmetries from various sources described in the text.}
\label{tab:systematics-table}
\begin{tabular}{lcccccccc} 
\hline
\hline
Source                 & $x$  & $y$  & $\Dx$ & $\Dy$ & Fit fraction & $\CalACP$ & Signal yield & Signal asymmetry \\
\hline
\BB\ yields            & 0.02 & 0.04 & 0.02 & ...  & 0.02 & 0.08  & 1.4  & 0.01 \\
\BB\ PDF               & ...  & ...  & ...  & ...  & ...  & 0.01  & 3.3  & ... \\
Signal PDF             & 0.02 & 0.02 & 0.01 & 0.02 & 0.01 & 0.08  & 48.3 & 0.01 \\
\qqbar\ Dalitz plot    & 0.06 & 0.03 & 0.05 & 0.05 & 0.03 & 0.14  & 47.7 & ... \\
\BB\ Dalitz plot       & 0.03 & 0.03 & 0.02 & 0.02 & 0.02 & 0.10  & 31.6 & 0.02 \\
Efficiency Dalitz plot & ...  & 0.01 & ...  & ...  & ...  & 0.03  & 0.6  & ... \\
Fit bias               & 0.01 & 0.02 & 0.02 & 0.01 & ...  & 0.05  & 2.8  & ... \\
\hline
\hline
\end{tabular}
\end{table*}

\renewcommand{\arraystretch}{1}

The fixed parameters of the signal \mes\ and \DeltaE\ PDFs are studied in
the control sample \BptoDzbpip; \DzbtoKpi. The parameters are determined
from data and MC samples, from which shift and scale factors are
calculated and used to adjust the parameters for the nominal fit. The
parameters are then varied in accordance with the errors on these shift and
scale factors and the fits are repeated.
Uncertainties due to the \mes\ distribution for the \qqbar\ background, which is    
fixed in the fit, are assessed to be negligible. The fit confirms the value of the
single parameter of the ARGUS function that is taken as input.

To confirm the fitting procedure, we perform 500 MC experiments in
which the events are generated from the PDFs used in the fit to data.
We repeat the exercise with \qqbar\ events alone drawn from the PDF
into which we embed signal and \BB\ background events randomly extracted
from the MC samples. 
Small fit biases are observed for some of the fit parameters and are
included in the systematic uncertainties.

Relative uncertainties in the efficiency, due to PID and tracking
efficiency corrections are $4.2\%$ and $2.4\%$ respectively; while \nbb\ has an
associated error of $1.1\%$.
The efficiency correction due to the selection requirement on the NN output has
also been calculated from \BptoDzbpip; \DzbtoKpi\ data and MC samples, and
is found to be $(96.2\pm1.2)\%$. The error on this correction is
incorporated into the branching fraction systematic uncertainties.

Measured \CP\ asymmetries could be affected by detector charge bias.
We include a systematic uncertainty of 0.005 to account for this effect~\cite{Aubert:2008bj}.
Furthermore, some of our selection requirements, for example that on the NN
output, may induce an asymmetry.  We estimate the possible size of such an
effect as 0.020 based on the study of our control sample.  

In addition to the above systematic uncertainties we also
estimate uncertainties from two sources related to the signal
Dalitz plot model.
The first of these elements consists of the parameters of the various
components of the signal model: the masses and widths of all intermediate
resonances, the value of the parameter that characterizes the nonresonant
shape, and the value of the Blatt--Weisskopf barrier radius.
The associated uncertainties are evaluated by adjusting the parameters
within their experimental errors and refitting.
The second element is the uncertainty due to the composition of the signal
model. It reflects observed changes in the parameters of the components
when the data are fitted with the less significant \fIII\ component removed
from the model and with one of the states \omegaI, \fI, \chiczero, or
\chictwo\ added to the model.
The uncertainties from each of these elements are added in quadrature to
obtain the final model-dependence.

\section{Discussion and Summary}

Our results are shown in Tables~\ref{tab:summarytable} 
and~\ref{tab:conclusion-results-summary}.
The Dalitz plot is dominated by the $\rhoI$ resonance and
a nonresonant contribution which, as seen in other charmless three-body
hadronic $B$ decays, is well modeled with an exponential form-factor.
The measured branching fraction for the decay
$\Bpm\to\rhoIpipm$ agrees with the world-average value~\cite{Amsler:2008zz}
and is consistent with theoretical predictions based on QCD factorization
models~\cite{Li:2006jb} and SU(3) flavor symmetry~\cite{Chiang:2008zb}.
The measured branching fraction of the nonresonant component is consistent
with some theoretical predictions~\cite{Cheng:2002qu,Deandrea:2000ce,Tandean:2002pe}. 
We find the parameter of the nonresonant lineshape to be 
$\alpha_{\rm nr} = 0.28\pm0.06\gevccSqinv$ (statistical uncertainties only),
which is comparable with values found in analyses of other charmless decay
modes such as 
$\B \rightarrow K\pi\pi$~\cite{Garmash:2004wa,Garmash:2005rv,Garmash:2006fh}  
and $B \rightarrow KKK$~\cite{Garmash:2004wa,Aubert:2006nu}.

Contributions from $\rhoII$ and $\fIII$ are also included in the Dalitz plot
model, where the mass and width of the \fIII\ are determined to be 
$m_{f_0(1370)} = 1400\pm40\mevcc$ and $\Gamma_{f_0(1370)} = 300\pm80\mev$ (statistical
uncertainties only).
We have made the first observation of the decay $\Bpm\to\fIIpipm$
with a statistical significance of $6.1\sigma$.
After correcting for the $\fII\to\pipi$ branching fraction of
$(84.8_{-1.2}^{+2.4}) \times 10^{-2} \times \frac{2}{3}$~\cite{Amsler:2008zz}, 
we obtain
\begin{eqnarray}
  {\cal B}(\Bpm\to\fIIpipm) & = & \\
  && \hspace{-20mm}
  (1.57 \pm 0.42 \pm 0.16 \,^{+0.53}_{-0.19}) \times 10^{-6}
  \nonumber
\end{eqnarray}
where the uncertainties are statistical, systematic, and model-dependent, respectively. 
The latter includes the uncertainty of the $\fII\to\pipi$ branching fraction.
The above measurements of the branching fractions are generally improved from
previous results~\cite{Aubert:2005sk}, although some of the uncertainties are
not reduced, largely due to a more realistic assignment of model-dependent
uncertainties in this analysis.

The $90\%$ confidence-level upper limits for the branching fractions of
$\Bpm\to\chiczero\pipm$ and $\Bpm\to\chictwo\pipm$ are found to be
\begin{eqnarray}
  {\cal B}(\Bpm\to\chiczero\pipm) & < & 1.5 \times 10^{-5} \\
  {\cal B}(\Bpm\to\chictwo\pipm) & < & 2.0 \times 10^{-5}
\end{eqnarray}
where the $\chi_{c(0,2)}\to\pipi$ widths are determined from recent
measurements by Belle~\cite{Nakazawa:2004gu}.
The absence of these charmonium contributions precludes the extraction of
the unitarity triangle angle $\gamma$ that has been proposed in the
literature~\cite{Eilam:1995nz,Deshpande:1995nu,Bediaga:1998ma,Bajc:1998bs,Deandrea:2000zg,Blanco:2000gw}. 

We do not find any signal for the decay $\Bpm\to\fIpipm$.  The branching
fraction upper limit we obtain is consistent with the prediction of a recent 
perturbative QCD calculation if the $\fI$ meson is dominated by an
$s\bar{s}$ component~\cite{Zhang:2008sa}.

We do not find any statistically significant $\CP$ asymmetries for the
components in the nominal Dalitz plot model.
The $\CP$ asymmetry in $\Bpm\to\rhoIpipm$ has a dependence on the
presence or absence of the \fIII\ term in the model.
The $\CP$ asymmetry of the \fIII\ term itself appears highly sensitive to the 
Dalitz plot model, 
varying dramatically between the favored and the second solution.
Since the presence of this component is not established, especially given
its insignificant contribution in the second solution, we set a $90\%$ CL
upper limit on its product branching fraction at $4.0 \times 10^{-6}$.

In conclusion, we have performed a Dalitz plot analysis of 
\BtoPPP\ decays based on a \onreslumi\ data sample containing
\bbpairs\ \BB\ pairs collected with the \babar\ detector.
Our model includes a momentum-dependent nonresonant component and four
intermediate resonance states: \rhoIpipm, \rhoIIpipm, \fIIpipm, and \fIIIpipm.
We do not find any significant contributions from 
\fIpipm, \chiczpipm, or \chictwopipm.
We find no evidence for direct \CP\ violation.
Our results will be useful to reduce model uncertainties in the extraction of
the CKM angle $\alpha$ from time-dependent Dalitz plot analysis of
$\Bz\to\pip\pim\piz$. The results presented here supersede those in our previous 
publication~\cite{Aubert:2005sk}.

\section{Acknowledgments}

We are grateful for the 
extraordinary contributions of our \pep2\ colleagues in
achieving the excellent luminosity and machine conditions
that have made this work possible.
The success of this project also relies critically on the 
expertise and dedication of the computing organizations that 
support \babar.
The collaborating institutions wish to thank 
SLAC for its support and the kind hospitality extended to them. 
This work is supported by the
US Department of Energy
and National Science Foundation, the
Natural Sciences and Engineering Research Council (Canada),
the Commissariat \`a l'Energie Atomique and
Institut National de Physique Nucl\'eaire et de Physique des Particules
(France), the
Bundesministerium f\"ur Bildung und Forschung and
Deutsche Forschungsgemeinschaft
(Germany), the
Istituto Nazionale di Fisica Nucleare (Italy),
the Foundation for Fundamental Research on Matter (The Netherlands),
the Research Council of Norway, the
Ministry of Education and Science of the Russian Federation, 
Ministerio de Educaci\'on y Ciencia (Spain), and the
Science and Technology Facilities Council (United Kingdom).
Individuals have received support from 
the Marie-Curie IEF program (European Union) and
the A. P. Sloan Foundation.

\bibliographystyle{apsrev}

\appendix

\section{Correlations Between Fit Fractions and Direct \CP\ Asymmetries}
\label{sec:correlations}

In \tabref{correlations} we present the statistical linear correlations
between the values of ${\it FF}_j$ and $\CalACP_{\!,\,j}$.

\renewcommand{\arraystretch}{1.5}

\begin{table*}[!htb]
\caption{Matrix of statistical correlation coefficients between fit fractions and direct
  \CP\ asymmetries.}
\label{tab:correlations}
\begin{center}
\begin{tabular}{lrrrrrrrrrr}
\hline
\hline
                &\tcc{\rhoI}              & \tcc{\rhoII}            & \tcc{\fII}              & \tcc{\fIII}        & \tcc{Nonresonant} \\
Parameter       & \occ{$FF$} & \occ{$\CalACP$} & \occ{$FF$} & \occ{$\CalACP$} & \occ{$FF$} & \occ{$\CalACP$} & \occ{$FF$} & \occ{$\CalACP$} & \occ{$FF$} & \occ{$\CalACP$} \\
\hline
$\rhoI$ $FF$       &  $1.00$ \\
$\rhoI$ $\CalACP$  &  $0.00$ &    $1.00$   \\
$\rhoII$ $FF$      &  $0.10$ &    $0.29$   &  $1.00$ \\
$\rhoII$ $\CalACP$ &  $0.35$ &    $0.14$   &  $0.31$ &    $1.00$   \\
$\fII$ $FF$        & $-0.14$ &   $-0.02$   & $-0.15$ &   $-0.05$   &  $1.00$ \\
$\fII$ $\CalACP$   & $-0.12$ &    $0.06$   & $-0.01$ &   $-0.02$   & $-0.08$ &    $1.00$   \\
$\fIII$ $FF$       & $-0.28$ &   $-0.16$   & $-0.45$ &   $-0.31$   & $-0.09$ &    $0.02$   &  $1.00$ & \\
$\fIII$ $\CalACP$  & $-0.14$ &    $0.00$   & $-0.09$ &   $-0.13$   &  $0.01$ &   $-0.10$   & $-0.14$ &   $1.00$ \\
Nonresonant $FF$   & $-0.50$ &    $0.06$   & $-0.15$ &    $0.06$   & $-0.17$ &    $0.07$ &  $0.26$ & $-0.17$  & $1.00$ \\
Nonresonant $\CalACP$   & $-0.02$ &   $-0.22$   &  $0.03$ &    $0.09$   &  $0.03$ &   $-0.06$   &  $0.13$ &    $0.27$  & $0.06$ & $1.00$ \\
\hline
\hline
\end{tabular}
\end{center}
\end{table*}

\section{Comparison of Results in Favored and Second Solution}
\label{sec:secondSoln}

In \tabref{secondSoln} we give a comparison of the results for the two solutions. 
Note that the $\CalACP$ value of $\fIIIpipm$ is at the physical boundary in the
second solution.

\renewcommand{\arraystretch}{1.5}

\begin{table*}[hbt!]
\caption{Comparison of results for the favored and second solutions.
The uncertainties are statistical only.}
  \label{tab:secondSoln}
\begin{tabular}{lr@{$\pm$}lr@{$\pm$}l|r@{$\pm$}lr@{$\pm$}l}
  \hline
  \hline
  & \multicolumn{4}{c|}{Favored solution} 
  & \multicolumn{4}{c}{Second solution} \\
  \hline
  Inclusive signal yield   & \multicolumn{4}{c|}{$ 1219   \pm 50    $} & \multicolumn{4}{c}{$ 1195   \pm 45    $} \\
  Inclusive signal \CalACP\ & \multicolumn{4}{c|}{$ +0.032 \pm 0.044 $} & \multicolumn{4}{c}{$ +0.015 \pm 0.043 $} \\
  \qqbar\ background yield & \multicolumn{4}{c|}{$ 2337   \pm 62    $} & \multicolumn{4}{c}{$ 2358   \pm 64    $} \\
  \qqbar\ background \CalACP\ & \multicolumn{4}{c|}{$ +0.002 \pm 0.027 $} & \multicolumn{4}{c}{$ +0.011 \pm 0.027 $} \\
  \hline
  & \multicolumn{4}{c|}{Favored solution} 
  & \multicolumn{4}{c}{Second solution} \\
  Resonance    & \tcc{Fit fraction}      & \multicolumn{2}{c|}{$\CalACP$}     & \tcc{Fit fraction}      & \tcc{$\CalACP$}          \\
  \hline
  $\rhoIpipm$  & $  0.532 $ & $  0.037 $ & $ +0.18 $ & $  0.07 $ \ & \ $  0.458 $ & $  0.033 $ & $ +0.03 $ & $  0.08 $ \\
  $\rhoIIpipm$ & $  0.091 $ & $  0.023 $ & $ -0.06 $ & $  0.28 $ & $  0.064 $ & $  0.016 $ & $ -0.54 $ & $  0.24 $ \\
  $\fIIpipm$   & $  0.059 $ & $  0.016 $ & $ +0.41 $ & $  0.25 $ & $  0.079 $ & $  0.016 $ & $ +0.55 $ & $  0.20 $ \\
  $\fIIIpipm$  & $  0.189 $ & $  0.033 $ & $ +0.72 $ & $  0.15 $ & $  0.030 $ & $  0.019 $ &  \tcc{$ -1.00 \,^{+0.58}_{-0.00}$} \\
  Nonresonant  & $  0.349 $ & $  0.042 $ & $ -0.14 $ & $  0.14 $ & $  0.365 $ & $  0.042 $ & $ -0.11 $ & $  0.14 $ \\
  \hline
  \hline
\end{tabular}
\end{table*}

\end{document}